\documentclass[11pt]{article}
\usepackage{amsmath}
\usepackage{mathrsfs,epsfig,morefloats}
\usepackage{amsfonts}
\usepackage{mathtools}
\usepackage{authblk}
\usepackage{xcolor}
\usepackage{graphicx}
\usepackage{float}
\usepackage{amsthm}
\usepackage{amssymb}
\usepackage{lscape}
\usepackage{rotating}
\usepackage{subfigure}
\usepackage{graphicx}

\usepackage{tabularx}
\usepackage{multirow}

\usepackage{algorithm}
\usepackage{apacite}

\allowdisplaybreaks

\usepackage{algpseudocode}

\usepackage{caption}
\usepackage{lipsum}

\DeclareCaptionFormat{algor}{%
  \hrulefill\par\offinterlineskip\vskip1pt%
    \textbf{#1#2}#3\offinterlineskip\hrulefill}
\DeclareCaptionStyle{algori}{singlelinecheck=off,format=algor,labelsep=space}
\usepackage{hyperref}
\hypersetup{
    colorlinks=true,
    linkcolor=blue,
    filecolor=blue,      
    urlcolor=blue,
citecolor=blue,
}
 \usepackage[sort&compress]{natbib}

\usepackage[section]{placeins}
\newcommand{\Date}[1]{\def\@Date{#1}}
\def\today{\number\day~\ifcase\month\or
 January\or February\or March\or April\or May\or June\or
 July\or August\or September\or October\or November\or December\fi~\number\year}

\def\be{\begin{equation}}
\def\ee{\end{equation}}
\def\bea{\begin{eqnarray}}
\def\eea{\end{eqnarray}}
\def\bd{\begin{displaymath}}
\def\ed{\end{displaymath}}
\def\bda{\begin{eqnarray*}}
\def\eda{\end{eqnarray*}}
\def\bsm{\begin{small}}
\def\esm{\end{small}}

\def\t0{\theta_0}

\def\ha1{\hat \beta_1}

\def\bnt{\begin{enumerate}}
\def\ent{\end{enumerate}}

\def\bsc{\begin{scriptsize}}
\def\esc{\end{scriptsize}}

\theoremstyle{definition}

\newcommand{\E}{\rm E}

\makeatletter
\newcommand{\figcaption}{\def\@captype{figure}\caption}
\newcommand{\tabcaption}{\def\@captype{table}\caption}
\makeatother

\usepackage{scalerel,stackengine}
\stackMath
\newcommand\reallywidehat[1]{%
\savestack{\tmpbox}{\stretchto{%
  \scaleto{%
    \scalerel*[\widthof{\ensuremath{#1}}]{\kern-.6pt\bigwedge\kern-.6pt}%
    {\rule[-\textheight/2]{1ex}{\textheight}}
  }{\textheight}%
}{0.5ex}}%
\stackon[1pt]{#1}{\tmpbox}%
}
\makeatletter
\newcommand{\printfnsymbol}[1]{%
  \textsuperscript{\@fnsymbol{#1}}%
}
\makeatother

\begin{document}

\title{MarZIC: A Marginal mediation model for Zero-Inflated Compositional mediators with applications to microbiome data}

\author[1]{Quran Wu}
\author[2]{James O'Malley}
\author[1]{Janaka S.S. Liyanage}
\author[1]{Susmita Datta}
\author[3]{Raad Z. Gharaibeh}
\author[3]{Christian Jobin}
\author[4]{Margaret R. Karagas}
\author[4]{Modupe O. Coker}
\author[4]{Anne G. Hoen}
\author[4]{Brock C. Christensen}
\author[4]{Juliette C. Madan}
\author[1]{Zhigang Li\thanks{Correspondence: Zhigang Li, Department of Biostatistics, University of Florida, Gainesville, FL 32611; Email: zhigang.li@ufl.edu.}}
\affil[1]{Department of Biostatistics, University of Florida,  Gainesville, FL, USA, 32611}
\affil[2]{The Dartmouth Institute, Geisel School of Medicine at Dartmouth, Hanover, NH, USA, 03755}
\affil[3]{Department of Medicine, University of Florida,  Gainesville, FL, USA, 32611}
\affil[4]{Department of Epidemiology, Geisel School of Medicine at Dartmouth, Hanover, NH, USA, 03755}

\date{}
\maketitle

\begin{abstract}
The human microbiome can contribute to pathogeneses of many complex diseases by mediating disease-leading causal pathways. However, standard mediation analysis methods are not adequate to analyze the microbiome as a mediator due to the excessive number of zero-valued sequencing reads in the data that is compounded by its compositional structure. The two main challenges raised by the zero-inflated data structure are: (a) disentangling the mediation effect induced by the point mass at zero; and (b) identifying the observed zero-valued data points that are actually not zero (i.e., false zeros). We develop a novel marginal mediation analysis method under the potential-outcomes framework to fill this gap and show the marginal model can also account for the compositional structure. The mediation effect can be decomposed into two components that are inherent to the two-part nature of zero-inflated distributions. With probabilistic models to account for observing zeros, we also address the challenge with false zeros. A comprehensive simulation study and the application in a real microbiome study showcase our approach in comparison with existing approaches.
\end{abstract}


\section{Introduction\label{sc:intro}}

Emerging evidence suggest that the human microbiome and the immune system are constantly shaping each other \citep{Belkaid2014}. Thus the human microbiome can contribute to disease pathogeneses by mediating disease-leading causal pathways in complex diseases such as Alzheimer's disease \citep{Wang2019a} and cancer \citep{Jin2019,Tanoue2019}. To study human microbiome, 16S ribosomal RNA gene sequencing and metagenomic shotgun sequencing have been popular methods to quantify microbiome composition in microbiome studies. A challenging feature of microbiome sequencing data is that it has excessive number of zeros \citep{hongzhe18review}. Many microbiome data sets have more than 50\% of the sequencing reads being 0, and it could be as high as $80\%$ or more. These zeros are likely to be a mixture of structural zeros (i.e., true zeros) that represent true absence of microbial taxa and undersampling zeros (i.e., false zeros) that result from failure of detection. The zero-inflated data feature compounded by a compositional structure poses a challenge that needs to be addressed specifically in mediation analyses. Although there have been some exciting efforts to model microbiome as a high-dimensional mediator \citep{Sohn2019,chanWang2019,LeiLiu2019}, it remains a daunting task to address the zero-inflated data structure.   

Mediation analysis is an important tool to investigate the role of intermediate variables (i.e., mediators) in a causal pathway where the causal effect partially or completely relies on the mediators. For example, people with higher socioeconomic status tend to have longer life expectancy, but this causal pathway may be explained by many possible mediators including access to better health care, fewer stressors, better living environment and so forth. In a mediation analysis, the indirect effect (i.e., mediation effect) through one or more mediators can be estimated and tested along with the direct effect. This technique was first popularized in psychology and social sciences and it has become a common analysis tool in many research areas such as epidemiology, environmental health sciences, medicine, randomized trials and psychiatry. There are two general types of mediation analysis approaches: potential-outcomes (PO) or counterfactual-outcomes methods \citep{Vander09,Imai,Vander15} and traditional linear mediation analysis methods \citep{BaronKenny,MacKin08}. The former approach stems from a counterfactual nonparametric function of a causal relationship without relying on linear assumptions and the latter is based on linear regression models. These approaches coincide with each other under linearity assumptions. PO approaches are more flexible because they can allow interaction effects of the independent variable with mediators as well as nonlinear effects. Reviews of mediation analysis approaches and their assumptions can be found in the literature \citep{MacKinnon2007,Vander16,Lange2017}. 

Although mediation modeling frameworks have been well established, to the best of our knowledge, there have been few studies to address zero-inflated compositional mediators. In a typical mediation analysis, the total effect of an independent variable can be decomposed into a mediation effect and a direct effect where the mediation effect measures the amount of the total causal effect attributable to change in the mediator caused by the independent variable and the direct effect measures the causal effect due to change in the independent variable while keeping the mediator variable constant. When the mediator has a marginal zero-inflated distribution such as a zero-inflated Beta (ZIB) distribution, we show that its mediation effect can be further decomposed into two parts with one part being the mediation effect attributable to the amount of numeric change in the mediator and the other part being the mediation effect attributable to the binary change of the mediator from zero to a non-zero state. This phenomenon can be explained by the two-part nature of a zero-inflated distribution. For example, a ZIB distribution is essentially a two-component mixture distribution \citep{Dalrymple2003}: one component is a degenerate distribution with probability mass of one at zero, and the other component is a Beta distribution. The mediator changing from zero to a positive value results in the discrete jump from zero to a non-zero state as well as the change in the numerical metric of the mediator and thus the mediation effect can be decomposed accordingly. Both changes have important interpretations in microbiome research. What makes it more complicated is that the observed zero-valued data points could be false zeros meaning that the true values are non-zero but observed as zero due to failure of detection. This is similar to a missing data problem and will be addressed here as well.

To fill the research gap in mediation modeling development, we propose a novel marginal mediation analysis approach under the PO framework to deal with zero-inflated compositional mediators. This approach can allow a mixture of truly zero-valued datapoints and false zeros. Our method is able to decompose the mediation effect into two components that are inherent to zero-inflated mediators: one component is the mediation effect attributable to the numeric change of the mediator on its continuum scale and the other component is the mediation effect attributable to the binary change of the mediator from zero to a non-zero state. So the mediation effect is actually the total mediation effect of the two components each of which can be estimated and tested. An extensive simulation study is conducted to evaluate our approach MarZIC in comparison with a standard PO mediation analysis approach \citep{Imai} and another approach \citep{Sohn2019} that can analyze microbiome composition as a mediator.  

We introduce the model and its associated notations in Section \ref{sc:model}. Estimation and inference procedures are provided in Section \ref{sc:estimation}. A simulation study to assess the performance of our model in comparison with existing approaches is presented in Section \ref{sc:simu}, followed by an application of our model in Section \ref{sc:app}, and a discussion in Section \ref{sc:discu}. Additional details and derivations can be found in the Appendix.

\section{Model and Notation}\label{sc:model}
For simplicity, we suppress subject index in all notations in this section. Let $Y$, $M=(M_1,\dots,M_{K+1})$ and $X$ denote the continuous outcome variable, the compositional mediator variable and the independent variable respectively. For example, $M$ could be the vector of relative abundances (RA) of microbial taxa. Before constructing the model for zero-inflated data, we first describe the model for the special case where the mediator $M$ have no zeros which could happen if investigators choose to impute zeros with a Pseudocount or a small positive number. The model for zero-inflated data will be provided after that.

\subsection{Model for data without zeros}\label{sc:modelNo0}
In this subsection, we assume there are no zeros for the mediator $M$ in the data which is very rare, but it could happen if zeros are replaced by a Pseudocount or a small positive number. Let $M$ follow a $(K+1)-$dimensional Dirichlet distribution indexed by its mean parameters $\mu_1,\dots,\mu_{K+1}$ with $\sum_{k=1}^{K+1}\mu_k=1$ and a dispersion parameter $\phi$. We assume the outcome $Y$ depends on $M$ and $X$ through the following regression equation: 
\begin{align}
Y=\sum_{k=1}^{K+1}\beta^kM_k+\beta^XX+\sum_{k=1}^{K+1}\beta^{kk}XM_k+\epsilon \label{ge:1no0}
\end{align}
where the random error $\epsilon$ follows a normal distribution with mean of 0 and a constant variance, $\beta^k$, $\beta^X$ and $\beta^{kk}$ are regression coefficients, and $XM_k$ is the interaction term between the independent variable $X$ and the mediator $M_k$. All taxa and their interactions with $X$ are included in the model, and thus the compositional structure is accounted for in this model. Later, we will show that a marginal model can also  account for the compositional structure. Equation (\ref{ge:1no0}) implies that the marginal association between $Y$ and any taxon $M_j, \hspace{0.1cm}j=1,\dots,K+1$, has the following form (derivation can be found in the Appendix):
\begin{align}
E_X(Y|M_j)=\beta_0^*+\beta_1^*M_j+\beta_2^*X+\beta_3^*XM_j, \label{ge:marginAssoNo0}
\end{align}
where $E_X(Y|M_j)$ is the mean of $Y$ conditional on $M_j$ given $X$, and
\begin{align*}
\beta_0^*=\frac{\sum_{k\ne j}\beta^k\mu_k}{\sum_{l\ne j}\mu_l},\hspace{0.2cm} \beta_1^*=\beta^j-\beta_0^*,\hspace{0.2cm}
\beta_2^*=\beta^X+\frac{\sum_{k\ne j}\beta^{kk}\mu_k}{\sum_{l\ne j}\mu_l},\hspace{0.2cm}\beta_3^*=\beta^{jj}-\frac{\sum_{k\ne j}\beta^{kk}\mu_k}{\sum_{l\ne j}\mu_l}.
\end{align*}
Therefore, without violating model (\ref{ge:1no0}), we can construct the following marginal regression model for the association between $Y$ and $M_j$ and $X$ such that it is equivalent to model (\ref{ge:1no0}):
\begin{align}
Y=\beta_0+\beta_1M_j+\beta_2X+\beta_3XM_j+\epsilon^*, \label{ge:marginModNo0}
\end{align}
where the random error $\epsilon^*$ has a normal distribution with mean of 0. An advantage of the above marginal model over model (\ref{ge:1no0}) is that it is straightforward to interpret the regression coefficient $\beta_1$ as a typical regession coefficient, whereas the corresponding regression coefficient $\beta^j$ in equation (\ref{ge:1no0}) does not have such a straightforward interpretation. That is because there has to be at least one $M_k,\hspace{0.1cm}k\ne j$, changing when $M_j$ changes due to the compositional structure, and thus it is not possible to hold all $M_k$'s, $k\ne j$, constant while changing $M_j$ to interpret $\beta^j$ as a typical regession coefficient.

Another nice feature of marginal model (\ref{ge:marginModNo0}) is that the true values of its regession parameters ($\beta_0$, $\beta_1$, $\beta_2$ and $\beta_3$) are functions of the parameters $\mu_1,\dots,\mu_{K+1}$ of the Dirichlet distribution of $M$ as shown in equation (\ref{ge:marginAssoNo0}); therefore, the marginal model accounts for the compositional structure. 

It is also much more convenient to work on the marginal model (\ref{ge:marginModNo0}) due to its simpler form. With that and the above advantages, we propose to use the marginal model (\ref{ge:marginModNo0}) for constructing the mediation model. Under the Dirichlet distribution for $M$, the marginal distribution of $M_j$ is a Beta distribution with mean paramer $\mu_j$ and scale parameter $\phi$. The following equation can be used to model the association between $M_j$ and $X$:
\begin{align}
\ln{\Big(\frac{\mu_j}{1-\mu_j}\Big)}=\alpha^0+\alpha^1X. \label{ge:2no0}
\end{align}
Equations (\ref{ge:marginModNo0}) and (\ref{ge:2no0}) together form our marginal mediation model for the scenario without zeros for $M$.

\subsection{Model for data with zeros}
Now we consider scenarios where the data for $M$ contain zeros. Given the advantages of a marginal model as demonstrated in the above subsection, we will again use a marginal model for the association between $Y$ and any taxon $M_j$ to form a mediation model. For any taxon $M_j$, we construct the marginal model as follows: 
\begin{align}
Y=\beta_0+\beta_1M_j+\beta_2 1_{(M_j>0)}+\beta_3X+\beta_4X1_{(M_j>0)}+\beta_5XM_j+\epsilon \label{ge:1}
\end{align}
where $1_{(\cdot)}$ is an indicator function indicating whether $M_j$ is 0, the random error $\epsilon$ follows a normal distribution $N(0,\delta)$, and $\beta_1$, $\beta_2$, $\beta_3$, $\beta_4$ and $\beta_5$ are regression coefficients. An advantage of using $M_j$ instead of $\ln{(M_j)}$ in the model is that it does not require imputing zeros with a positive number. This model is fully compatible with allowing interactions between the independent variable and mediators as the two interaction terms: $X1_{(M_j>0)}$ and $XM_j$ are included in equation (\ref{ge:1}). In practice, investigators can also include only one or no interaction term depending on the hypothesis of interest.

For the marginal distribution of $M_j$, it is natural to use a zero-inflated Beta (ZIB) distribution because the marginal of a Dirichlet distribution is a Beta distribution \citep{Chen2016,Chai2018}. Its two-part density function is given as follows:
\begin{align*}
  f(m)=\begin{cases}
    \Delta, & m=0\\
    (1-\Delta)\frac{m^{\mu_j\phi-1}(1-m)^{(1-\mu_j)\phi-1}}{B\big(\mu_j\phi,(1-\mu_j)\phi\big)}, &m>0
  \end{cases}
\end{align*}
where $\Delta$ is the probability of being 0, $B(\cdot,\cdot)$ is the Beta function and $\mu_j$ and $\phi$ are the mean and dispersion parameters respectively of the Beta distribution for the non-zero part \citep{betaDist,betaReg}. To model the association of the mediator $M_j$ with $X$, we use the following equations:
\begin{align}
&\ln{\Big(\frac{\mu_j}{1-\mu_j}\Big)}=\alpha_0+\alpha_1X, \label{ge:21}\\
&\ln\bigg(\frac{\Delta}{1-\Delta}\bigg)=\gamma_0+\gamma_1X. \label{ge:22}
\end{align}

Equations (\ref{ge:1})-(\ref{ge:22}) together form our mediation model. The parameter $\alpha_1$ in equation (\ref{ge:21}) measures the association between $X$ and the RA level of the mediator and $\gamma_1$ in equation (\ref{ge:22}) measures the association between $X$ and the binary presence of the mediator. Notice that $X$ is a scalar here, but it is obvious that other covariates such as potential confounders can be included in the model  equations.

\subsection{Mechanism for observing zeros of the mediator}
For microbiome abundance data, observations that cannot be detected are set to be zero. Consequently, there are two types of zeros in the observed abundance data: true abundance of zero (i.e., absence) and abundance that is reported as zero as a consequence of the measurement failure. We will use real microbiome studies to illustrate our method in a later section. Let $M_j^*$ denote the observed value of $M_j$. When the observed value is positive (i.e., $M_j^*>0$), we assume that $M_j^*=M_j$. But when $M_j^*=0$, we don't know whether $M_j$ is truly zero or $M_j$ is positive but observed as zero. We consider the following mechanism for observing a zero of the microbial taxon abundance: 
\begin{align}\label{eq:RAzero1}
\Pr(M_j^*=0|M_j,L)=1_{(M_jL<1)},
\end{align}
where $L$ is the library size (i.e., sequencing depth) and the product $M_jL$ can be interpreted as the sample absolute abundance (SAA) of the $j$th taxon in a sample. Under this mechanism, all SAA below 1 have an observed value of zero. Here 1 can be considered as the Limit of Detection (LOD). We refer to this mechanism as "LOD mechanism" hereafter. Since SAA depends on both $L$ and $M_j$, the LOD mechanism is not deterministic conditional on the library size. The probability of observing a zero conditional on $L$, the library size, is equal to $\E(1_{(M_jL<1)}|L)=\Pr(M_j<1/L)$.

\subsection{Marginal mediation effect and direct effect}\label{NIE_NDE}
Under the potential-outcomes (PO) framework \citep{Vander16}, we can define the natural indirect effect (NIE), natural direct effects (NDE) and controlled direct effect (CDE) where NIE is the mediation effect. We refer to NIE as the marginal mediation effect because the proposed mediation models are based on marginal models as shown in Section \ref{sc:model}. The total effect of $X$ is equal to the summation of NIE and NDE. Let $M_j(x)$ denote the value of $M_j$ if $X$ equals $x$. Let $Y_{xm}$ denote the value of $Y$ if $(X,M_j)=(x,m)$. The average NIE, NDE and CDE for $X$ changing from $x_1$ to $x_2$ are defined as:

$\text{NIE}=\E\big(Y_{x_2M_j(x_2)}-Y_{x_2M_j(x_1)}\big)$

$\text{NDE}=\E\big(Y_{x_2 M_j(x_1)}-Y_{x_1 M_j(x_1)}\big)$

$\text{CDE}=\E\big(Y_{x_2 m}-Y_{x_1 m}\big),\hspace{0.1cm}\text{for a fixed (i.e., controlled) value of}\hspace{0.1cm} M_j=m,$

\noindent where $Y_{x_2M_j(x_1)}$ is a counterfactual outcome. By plugging the equations (\ref{ge:1})-(\ref{ge:22}) into the above definitions and using Riemann-Stieljes integration \citep{stiel}, we can obtain the following formulas:
\begin{align*}
\text{NIE}&=\E(Y_{x_2 M_j(x_2)})-\E(Y_{x_2 M_j(x_1)})=\E(\E(Y_{x_2 M_j(x_2)}|M_j(x_2)))-\E(\E(Y_{x_2 M_j(x_1)}|M_j(x_1)))\\ 
&=\E(\beta_0+\beta_1M_j(x_2)+\beta_2 1_{(M_j(x_2)>0)}+\beta_3x_2+\beta_4x_21_{(M_j(x_2)>0)}+\beta_5x_2M_j(x_2))\\
&\hspace{0.5cm}-\E(\beta_0+\beta_1M_j(x_1)+\beta_2 1_{(M_j(x_1)>0)}+\beta_3x_2+\beta_4x_21_{(M_j(x_1)>0)}+\beta_5x_2M_j(x_1))\\
&=(\beta_1+\beta_5x_2)(\E(M_j(x_2))-\E(M_j(x_1)))+(\beta_2+\beta_4x_2)(\E(1_{(M_j(x_2)>0)})-\E(1_{(M_j(x_1)>0)}))\\
&=\text{NIE}_1+\text{NIE}_2,\\
\text{NIE}_1&=(\beta_1+\beta_5x_2)(\E(M_j(x_2))-\E(M_j(x_1)))\\
&=(\beta_1+\beta_5x_2)\Bigg(\int\limits_{m\in [0,1]}mdF_{M_j(x_2)}(m)-\int\limits_{m\in [0,1]}mdF_{M_j(x_1)}(m)\Bigg)\\
&=(\beta_1+\beta_5x_2)\Big(\text{expit}{(\alpha_0+\alpha_1x_2)}-\text{expit}{(\alpha_0+\alpha_1x_1)}\Big)\\
&-(\beta_1+\beta_5x_2)\Big(\text{expit}(\gamma_0+\gamma_1x_2)\text{expit}{(\alpha_0+\alpha_1x_2)}\\
&-\text{expit}(\gamma_0+\gamma_1 x_1)\text{expit}{(\alpha_0+\alpha_1x_1)}\Big),\\
\text{NIE}_2&=(\beta_2+\beta_4x_2)\big(\text{expit}(\gamma_0+\gamma_1 x_1)-\text{expit}(\gamma_0+\gamma_1 x_2)\big),
\end{align*}
where $\text{expit}(\cdot)$ is the inverse function of $\text{logit}(\cdot)$, $F_{M_j(x)}(m)$ denotes the CDF of $M_j(x)$ and $dF_{M_j(x)}(m)$ denotes the stieltjes integration \citep{stiel} with respect to $F_{M_j(x)}(m)$. So NIE, $\text{NIE}_1$, $\text{NIE}_2$, NDE and CDE can be estimated by plugging the parameter estimates into the formulas. Confidence intervals (CI) are obtained using the multivariate delta method as outlined in the Appendix. An alternative approach for finding standard errors to construct CI is bootstrapping \citep{efron1986}. $\text{NIE}_1$ can be interpreted as the marginal mediation effect due to the change of the mediator on its numeric scale and $\text{NIE}_2$ can be interpreted as the marginal mediation effect due to the discrete binary change of the mediator from zero to a non-zero status. This decomposition can be also seen in Figure 1 where there are two possible indirect causal pathways from $X$ to $Y$ through the mediator $M_j$. 

\begin{figure}[ht]
  \begin{center}
  \includegraphics[width =1\textwidth,angle=0]{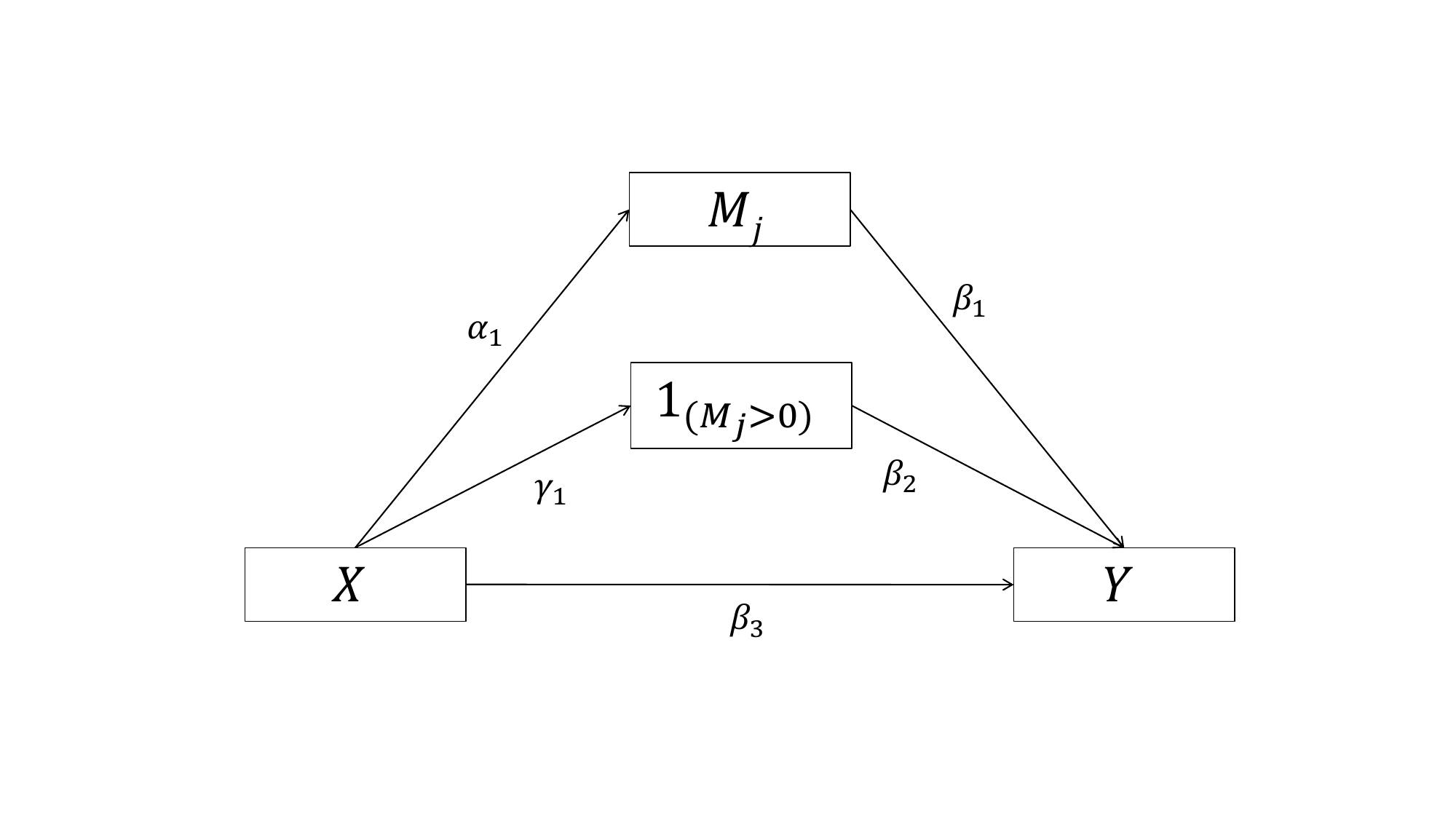}
  \end{center}
  \caption{Potential causal mediation pathways of a zero-inflated mediator.}
  \label{fig_MediatPath}
\end{figure}

\section{Parameter estimation}\label{sc:estimation}
Maximum likelihood estimation (MLE) will be used to estimate the parameters. The data that is needed to estimate the marginal mediation effects for the $j$th taxon is $(Y,R,M_j^*,L,X)$ where $R=1_{(M_j^*>0)}$. The estimation challenge is that $M_j$ is not always observable due to false zeros. The log-likelihood contribution from those subjects with false zeros cannot be directly calculated. However, given that we know the probability of observing a zero in equation (\ref{eq:RAzero1}), we can still obtain their log-likelihood contributions by integrating the joint density function over all possible values of $M_j$ using Riemann–Stieltjes integration \citep{stiel}. Let $(y_i,r_i,m_i^*,l_i,x_i)$ denote the observed data values of $(Y,R,M_j^*,L,X)$ for the $i$th subject in a study and $m_i$ denote the true value of the mediator $M_j$ for the $i$th subject. We use $i$ as subject index hereafter throughout the paper. The subjects can be divided into two groups by whether $m_i^*$ is non-zero and we derive the log-likelihood contribution for each group. The first group consists of subjects whose observed value of the mediator is non-zero (i.e., $m_i^*>0$). Based on the assumptions in the equations (\ref{ge:1})-(\ref{ge:22}) where $\epsilon$ is assumed to have a normal distribution, the log-likelihood contribution from the $i$th subject (if it is in group 1) can be calculated as:
\begin{align*}
&\ell_i^1=\ln(f(y_i, r_i|m_i^*,x_i,l_i)f(m_i^*|x_i,l_i))=\ln(f(y_i|m_i^*,x_i,l_i)p(r_i|m_i^*,x_i,l_i)f(m_i^*|x_i,l_i))\\
&=\ln(f(y_i|m_i^*,x_i,l_i))+\ln(p(r_i|m_i^*,l_i))+\ln(f(m_i^*|x_i,l_i))\\
&=-0.5\ln(2\pi)-\ln(\delta)-\frac{\big(y_i-\beta_0-\beta_1m_i^*-\beta_2-(\beta_3+\beta_4)x_i-\beta_5x_im_i^*\big)^2}{2\delta^2}\\
&\hspace{0.4cm}+\ln (1-\Delta_i)-\ln{\bigg(B\big(\mu_i\phi,(1-\mu_i)\phi\big)\bigg)}\\
&\hspace{0.4cm}+(\mu_i\phi-1)\ln{(m_i^*)}+\big((1-\mu_i)\phi-1\big)\ln{(1-m_i^*)},\label{betaG1}
\end{align*}
where $f(\cdot|m_i^*,x_i,l_i)$, $p(\cdot|m_i^*,x_i,l_i)$ and $f(\cdot|x_i,l_i)$ are the (conditional) density (or probability mass function) for $Y$, $R$ and $M_j$ respectively, $\Delta_i=\text{expit}(\gamma_0+\gamma_1 x_i)$ and $\mu_i=\text{expit}{(\alpha_0+\alpha_1x_i)}$. Let $F(m|x)$ denote the (conditional) cumulative distribution function for $M_j$. The second group consists of subjects with $m_i^*=0$. The log-likelihood contribution from the $i$th subject (if it is in group 2) can be calculated as:
\begin{align*}
&\ell_i^2=\ln(f(y_i,r_i,m_i^*|x_i))=\ln\bigg(\int\limits_{m\in [0, 1]} f(y_i|m,x_i) p(r_i|m)dF(m|x_i)\bigg)\\
&=\ln\Bigg(\frac{\Delta_i}{\sqrt{2\pi\delta^2}}\exp{\bigg(-\frac{(y_i-\beta_0-\beta_3x_i)^2}{2\delta^2}\bigg)}\\
&\hspace{0.4cm}+\int\limits_0^{1/l_i} f(y_i|m,x_i)(1-\Delta_i)\frac{m^{\mu_i\phi-1}(1-m)^{(1-\mu_i)\phi-1}}{B\big(\mu_i\phi,(1-\mu_i)\phi\big)}dm\Bigg)\\
&=-0.5\ln(2\pi)-\ln (\delta)+\ln\Bigg(
\Delta_i\exp{\bigg(-\frac{(y_i-\beta_0-\beta_3x_i)^2}{2\delta^2}\bigg)}+\frac{1-\Delta_i}{B\big(\mu_i\phi,(1-\mu_i)\phi\big)}\int\limits_0^{1/l_i}h_i(m)dm\Bigg)
\end{align*}
where
\begin{align*}
h_i(m)=&m^{\mu_i\phi-1}(1-m)^{(1-\mu_i)\phi-1}\\
&\times\exp\bigg(-\frac{\big(y_i-\beta_0-\beta_1m-\beta_2-(\beta_3+\beta_4)x_i-\beta_5x_im\big)^2}{2\delta^2}\bigg).
\end{align*}
Taken together, we have the complete log-likelihood function given by:
\begin{equation}\label{logL}
\ell=\sum_{i\in\text{group 1}}\ell_i^1+\sum_{i\in\text{group 2}}\ell_i^2.
\end{equation}
The MLE of the parameters can be obtained by maximizing the above complete log-likelihood function. With the parameter estimates and the observed Fisher information matrix, we will be able to calculate NIE, $\text{NIE}_1$, $\text{NIE}_2$, NDE and CDE and their CI's.

\section{Simulation}\label{sc:simu}
Extensive simulations were carried out to demonstrate the performance of our approach MarZIC in comparison with two existing approaches under two settings. In setting 1 where the mediator was generated by univariate ZIB distributions, we compared MarZIC with a current standard practice in causal mediation analyses developed by Imai, Keele and Tingley \citep{Imai} (IKT approach hereafter) which is a PO approach and can be implemented in R using the package ``mediation'' \citep{mediationRpackage}. The Marginal Structural Models \citep{Vander09} is also a standard PO approach with a very similar definition of indirect effect. These causal mediation analysis approaches were not developed to analyze microbiome data, and thus could have poor performance when applied to microbiome data. In setting 2 where the mediator was generated by multivariate zero-inflated Dirichlet distributions, MarZIC was compared with IKT and CCMM \citep{Sohn2019} which was developed specifically to model microbiome composition as a mediator. In all simulation settings, the independent variable $X$ was binary and generated using the Bernoulli distribution Ber(0.5) such that the number of subjects was balanced between the two groups. The LOD mechanism in equation (\ref{eq:RAzero1}) for observing zero-valued data points of the mediator was used to generate zeros for the mediator $M_j$. 

To mimic the real study data, the library size was generated by randomly picking the library size with replacement from the real study data in Section \ref{sc:app} where the library size ranges from 31,607 to 911,652. The RA data was generated in a way such that it mimicked the distribution of RA in the real data. We generated 100 random datasets for each of the simulation settings. Multivariate delta method was used to derive confidence intervals in all settings.

\subsection{Simulation setting 1}
In this setting, the outcome $Y$ was assumed to be a continuous variable and generated using equation (\ref{ge:1}) where $\beta_5$ is set to be 0 in the simulation and other true parameter values can be found in Table \ref{table:withIKT}. Similar to simulation studies in the literature \citep{Chen2016,Chai2018} where RA were generated individually, we generated individual taxon RA with ZIB distributions based on equations (\ref{ge:21})-(\ref{ge:22}). The sample size was 100 in each of the 100 random datasets. Two scenarios were considered for the taxon RA: low RA (Scenario 1: mean of positive RA is equal to 0.0025) and high RA (Scenario 2: mean of positive RA is equal to 0.5). About 20\% of all sequencing reads were generated as true zeros (i.e., structured zeros) in both scenarios. Under the LOD mechanism in equation (\ref{eq:RAzero1}), about 30\% sequencing reads were false zeros in Scenario 1 and there were no false zeros in Scenario 2 because the RA in Scenario 2 was high and thus SAA were greater than 1 for all truly non-zero RA. Model performance was evaluated by estimation bias, standard error, coverage probability (CP) of 95\% CI of the estimators for parameters and the mediation effects in this comparison. For Scenario 1, the simulation results (Table \ref{table:withIKT}) showed good performance for MarZIC in terms of bias and CP of the mediation effects and the parameter estimates. All the biases were small and the CP were around the desired level of 95\%. The IKT approach, however, had a poor performance with a large bias (84.81\%) and a small CP (9\%). These poor performances were likely due to the false zeros not being appropriately accounted for by the IKT approach. Another disadvantage of IKT is that it cannot decompose the mediation effect into $\text{NIE}_1$ and $\text{NIE}_2$. For Scenario 2 with high RA where there were no false zeros, MarZIC showed good performance again in terms of the performance measures. IKT also showed satisfactory performance for the estimation of the NIE  because there were no false zeros in the data under this scenario, but IKT cannot decompose the mediation effect according to the zero-inflated distribution of mediator.

\begin{table}[ht!]
\caption{Simulation results for comparison between MarZIC and IKT with sample size of $n=100$. Bias, percentage of the bias, the empirical standard errors, the the mean of estimated standard errors and the empirical coverage probability of the $95\%$ CI for each estimator is respectively reported under the columns Bias, Bias \%, SE, Mean SE and CP(\%). Mediation effects from the IKT approach are provided at the bottom part of the table.}
\label{table:withIKT}
\centering
\resizebox{\linewidth}{!}{
\begin{tabular}{c c c c c c cc cc c c c c cc} 
 \hline
&&\multicolumn{7}{c}{Low relative abundance (mean=0.0025)}&\multicolumn{7}{c}{High relative abundance (mean=0.5)} \\
\cline{2-8} \cline{10-16} 
 Parameter  & True & Mean&Bias & Bias &SE&Mean &CP(\%) &&True&Mean&Bias & Bias &SE&Mean &CP(\%)\\
/Effect&&Estimate&&\%&& SE&&&&Estimate&&\%&& SE&\\
 \hline\hline\\
\multicolumn{16}{c}{MarZIC}\\
NIE$_1$                      &0.10&0.11&0.01&	 10.0&0.08	&	0.07&	91
                           && 9.30& 9.11&-0.18&-1.98&2.68&	2.70&	96\\
NIE$_2$                      &0.55&	0.52&-0.03&	-5.67&0.55	&	0.56&	97
  				 &&0.55&0.50&-0.06&-10.15&	0.62&	0.56&	94\\
NIE                        &0.65&0.63&-0.02&-3.31&0.58	& 0.58&	96
                           &&9.85&9.61&-0.24&-2.44&	3.25&	3.20&	95\\
$\beta_0$            &-2.00&	-2.05&	-0.05&	-2.45&	0.32&	0.33&	96
                        &&-2.00&-1.92&	0.07&	3.82&	0.32&	0.29	&94\\ 
$\beta_1$            &100.00&	101.89&	1.89&   1.89&	18.04&	19.04&	97
                        &&100.00&99.96&	-0.04&	-0.04&	1.89&	1.74&	91\\
$\beta_2$            &4.00&4.05&	0.05&	1.37&	0.38&	0.36&94
                        &&4.00&3.93&	-0.07	&-1.73&	0.58&	0.57&	91\\
$ \beta_3$           &5.00&	5.08&	0.08&	1.53&	0.53& 0.51&	94
                        &&5.00&4.97&	-0.03&	-0.62&	0.46&	0.46&	99\\
$ \beta_4$           &3.00&	2.93&	-0.07&	-2.40&	0.58& 0.55&	92
                        &&3.00&3.02&	0.02&	0.55&	0.53&	0.54&	99\\
$ \delta$              &1.00&	0.99&	-0.01&	-1.00&	0.07& 0.07&	90
                        &&1.00&0.97&	-0.03&	-2.99&	0.07&	0.07&	89\\
$\alpha_0$           &-6.20&	-6.24&	-0.04&-0.69&	0.36& 0.36&	94
                        &&-1.00&-1.01&	-0.01&	-0.93&	0.05&	0.05&	90\\
$\alpha_1$           &0.40&0.42&	0.02& 5.52&	 0.33& 0.29&	92
                        &&0.40&0.41&0.01&	1.69&	0.06&	0.07	&95\\ 
$ \xi$                   &50.00&56.42&	6.42&	12.83&	24.21&	19.35&97
                        &&50.00&53.37&	3.37&	6.74&	8.22&	8.40&	96\\
$ \gamma_0$       &-1.16&	-1.23&-0.07&	-5.75&	0.35&	0.36&99
                        &&-1.16&-1.20&	-0.04&	-3.18&	0.37	&0.34&	95\\
$\gamma_1$        &-0.50&	-0.53&	-0.03&	-5.10&	0.55&	0.55&	97
                        &&-0.50&-0.47&	0.03&6.91&	0.58&	0.53&	91\\
\multicolumn{16}{c}{IKT}\\
NIE            &0.65&	0.10&-0.55&	-84.81&	-&	-&	9
                 &&9.85&9.20&	-0.65&	-6.62&	-&	-&	94\\
\\[1ex] 
 \hline
\end{tabular}}
\end{table}

\subsection{Simulation setting 2}
In this setting, we generated microbiome RA data with multivariate zero-inflated Dirichlet distributions. Multiple testing was adjusted using the Benjamini-Hochberg Procedure \citep{Benjamini1995} in this setting such that the targeted FDR is 10\%. In this section, we suppressed the subject index $i$ in all notations for simplicity. 100 data sets were randomly generated for each case in this setting. As shown in Table \ref{table_simu_bin_cont_med}, six different cases were considered, of which some had sample size larger than the number of taxa and the others had sample size smaller than the number of taxa. Since CCMM needs to impute zero values with a positive number because it requires all RA to be non-zero in its analysis, we generated zero-valued data points for only the first taxon (to minimize the imputation burden for CCMM in the comparison) with equation (\ref{ge:22}). Let $K+1$ be the number of taxa. When the first taxon was zero, the rest of the taxa (i.e. taxon 2 to taxon K+1) was generated by the $K-$dimensional Dirichlet distribution with the mean parameter $(\mu_2,\mu_3,\dots,\mu_{K+1})^T$ and dispersion parameter $\phi$ where
\begin{align*}
\mu_k=\frac{\exp{(\alpha_0^k)}}{1+\sum_{k=2}^K\exp{(\alpha_0^k)}}, \hspace{0.2cm} k \in \{2,\dots,K \},\hspace{0.2cm} \text{and}\hspace{0.2cm}
\mu_{K+1}=\frac{1}{1+\sum_{k=2}^K\exp{(\alpha_0^k)}}.
\end{align*}
Notice that $\sum_{k=2}^{K+1}\mu_k=1$. When the first taxon was non-zero, the RA of all taxa was generated by the $(K+1)-$dimensional Dirichlet distribution with the mean parameter $(\mu_1^*,\mu_2^*,\mu_3,\dots,\mu_{K+1})^T$ and the dispersion parameter $\phi$ where $\mu_1^*=\frac{\mu_2\exp(a_0+a_1X)}{1+\exp(a_0+a_1X)}$ and $\mu_2^*=\mu_2-\mu_1^*$. After generating true RA, we then generate false zeros for the first taxon with LOD mechanism in (\ref{eq:RAzero1}) where library size was generated from the empirical distribution of library size in the real study data. $(\alpha_0^3,\dots,\alpha_0^K)$ were generated from uniform distribution $U(0,1)$. $a_0$ and $a_1$ were set to be -2 and 5 respectively. The percentage of false zeros for taxon 1 was set to be around 20\%. $\gamma_0$ and $\gamma_1$ were set to be 0 and -3 respectively so that the percentage of total zeros (including structural zeros and false zeros) was around 50\% in the data. The dispersion parameter $\phi=50$ to mimic overdispersion in real data. Notice that under this setting, only the means of the first taxon and second taxon were depending on $X$. The probability of absence of the first taxon depended on $X$ as well.

The outcome $Y$ was generated using the following equation:  
\begin{align}
Y=\beta_0+ \beta_{11}M_{1}+ \beta_{12}M_{2} + \beta_{2} 1_{(M_{1}>0)}+\beta_3X+  \beta_{4}X1_{(M_{1}>0)}+\beta_5XM_1+\epsilon. \label{scenario4Y}
\end{align}
where $M_1$ and $M_2$ denote the RA of the first taxon and the second taxon respectively, $(\beta_0,\beta_{11},\beta_{12},\linebreak\beta_2,\beta_3,\beta_4,\beta_5)=(4,90,10,2,1,1,1)$ and $\epsilon$ follows the standard normal distribution. In the data analysis step of the simulation, MarZIC analyzed each taxon as a mediator one by one whereas CCMM employed $\ell_1$ regularization to handle high dimensionality. For analyzing a taxon without any zeros, MarZIC used the model for data without zeros as described in Section \ref{sc:modelNo0}.

Notice that the data generation model (\ref{scenario4Y}) involves both $M_1$ and $M_2$. The relationships between $X$ and $\mu_1^*$ and $\mu_2^*$ are different from the data analysis model (\ref{ge:21}), so this simulation can also demonstrate the robustness of MarZIC with respect to model mis-specification to some extent. Under the data generation model (\ref{scenario4Y}), $Y$ has marginal associations with all taxa, but only the first two taxa marginally mediate the effect of $X$ on $Y$ because only their marginal mean values $\mu_1^*$ and $\mu_2^*$ depend on $X$ conditional on their presence. The indicator variable for the first taxon $1_{(M_1>0)}$ also has a mediation effect because the probability of its presence depends on $X$ since $\Delta=\text{expit}{(-3X)}$ for the simulated data. In summary, NIE$_1$ should be significant for $M_1$ and $M_2$, and NIE$_2$ should be significant for $M_1$ in the analysis results of this simulation.

Three indices were used to evaluate the model performance: Recall, Precision and F1 which were calculated as follows:
\begin{align*}
\text{Recall}=\frac{TP}{TP+FN}, \hspace{0.5cm}\text{Precision}=\frac{TP}{TP+FP},\hspace{0.5cm} \text{F1}=\frac{2}{\frac{1}{\text{recall}}+\frac{1}{\text{precision}}}
\end{align*}
where $TP$, $FP$, $TN$ and $FN$ denote true positive, false positive, true negative and false negative respectively. Recall is a measure of statistical power, the higher the better. Precision has an inverse relationship with false discovery rate (FDR) which is equal to (1-Precision), and thus the higher the Precision, the lower the FDR. When FP=0, Precision was set to be 1 regardless of whether TP=0. F1 is the Harmonic mean \citep{Hmean} of Recall and Precision that measures the overall performance in terms of Recall and Precision. The targeted FDR level is set to be 10\% for all the three approaches in this comparison which means that targeted Precision should be 90\%. 

The simulation results (See Table \ref{table_simu_bin_cont_med}) showed that MarZIC had a very good overall performance for identifying NIE$_1$ and NIE$_2$ in terms of Recall ($>$90\%), Precision ($>$90\%) and F1 ($>$90\%). MarZIC achieved the targeted Precision of 90\% across all cases. Precision was not applicable for NIE$_2$ in this setting because there was only one taxon having zero-valued sequencing reads in this simulation setting, and thus F1 was not applicable for NIE$_2$ either. CCMM had fair performance in terms of Recall (54.5-75.5\%), but its Precision rates (10.5-49.3\%) were much lower than the targeted Precision rate (90\%) which resulted in low F1 values (18.2-48.2\%). This suboptimal performance is likely due to (a) CCMM was proposed to model the RA on log-scale whereas equation (\ref{scenario4Y}) is on the original scale of RA, (b) CCMM was not developed to incorporate the mediation effect of the binary variable $1_{(M_1>0)}$ and (c) CCMM could not handle interactions between the independent variable and mediators such as $X1_{(M_1>0)}$ in model (\ref{scenario4Y}). CCMM could not generate any results for those cases with the number of taxa greater than or equal to 300 (See Table \ref{table_simu_bin_cont_med}) due to computational issues whereas MarZIC can handle all cases very well. This is likely because CCMM is too computationally demanding for its $\ell_1$ regularization algorithm which is not computationally capable of handling such high dimensionality. IKT had good Precision rates ($>$99.5\%), but comparably lower recall rate (53.5-59.5\%) compared to MarZIC, and thus also lower F1 rate.

\begin{table}[ht!]
\caption{Simulation results for the comparison of MarZIC with CCMM and IKT. Here $n$ denotes the sample size and $K+1$ denotes the number of taxa. (* Recall for NIE$_2$ is essentially the statistical power because only one taxon had zeros and was analyzed for estimating NIE$_2$.)}
\label{table_simu_bin_cont_med}
\centering
\resizebox{\linewidth}{!}{%
\begin{tabular}{c c c c c c cc cc c ccccc c c c c c } 
 \hline
&&\multicolumn{4}{c}{Recall* (\%)}&&\multicolumn{3}{c}{Precision (\%)} &&\multicolumn{3}{c}{F1 (\%)} \\
\cline{3-6} \cline{8-10} \cline{12-14} \noalign{\vskip .06in}
$K+1$ & $n$ &MarZIC&MarZIC& CCMM &IKT & &MarZIC& CCMM &IKT &&MarZIC& CCMM &IKT\\
&&(NIE$_1$)&(NIE$_2$)&&&&(NIE$_1$)&&&&(NIE$_1$)\\
\hline\hline\\

10	&200  & 99.50 & 88.00 & 54.50 & 56.50 && 99.00 & 49.30 & 100.00 && 99.10 & 48.20 & 71.00 \\
25    &200   & 99.50 & 84.00 & 63.00 & 59.50 && 99.30 & 27.90 & 100.00 && 99.30 & 36.80 & 73.00 \\ 
50	&200  & 99.00 & 95.00 & 63.00 & 56.50 && 97.00 & 13.70 & 99.50 && 97.50 & 22.20 & 70.80 \\
100	&200  & 98.50 & 92.00 & 75.50 & 53.50 && 96.80 & 10.50 & 100.00 && 97.10 & 18.20 & 68.70 \\ 
300    &200   & 97.00 & 91.90 & - & 55.00 && 98.50 & - & 99.50 && 97.10 & - & 69.50 \\ 
500	&200  & 99.00 & 91.00 & - & 56.50 && 99.20 & - & 100.00 && 98.80 & - & 70.70 \\ 

\\[1ex] 
 \hline
\end{tabular}}
\end{table}

\section{Real study application}\label{sc:app}
VSL\#3 is a commercially available probiotic cocktail (Sigma-Tau Pharmaceuticals, Inc.) of eight strains of lactic acid-producing bacteria: {\it Lactobacillus plantarum, Lactobacillus delbrueckii subsp. Bulgaricus, Lactobacillus paracasei, Lactobacillus acidophilus, Bifidobacterium breve, Bifidobacterium longum, Bifidobacterium infantis, and Streptococcus salivarius subsp}. Orally administered VSL\#3 has shown success in ameliorating symptoms and reducing inflammation in human pouchitis \citep{Gionchetti2000} and ulcerative colitis \citep{Sood2009}. Preventive VSL\#3 administration can also attenuate colitis in Il10-/- mice \citep{Madsen2001} and ileitis in SAMP1/YitFc mice \citep{Pagnini2010}. When used as a preventative strategy, it has the potential capability to prevent inflammation and carcinogenesis. In a mouse model, Arthur et al. \citep{Arthur2013} studied the ability of a probiotic cocktail VSL\#3 to alter the colonic microbiota and decrease inflammation-associated colorectal cancer when administered as interventional therapy after the onset of inflammation. The study duration was 24 weeks. In this study, there were 24 mice of which 10 were treated with VSL\#3 and 14 served as control. Gut microbiome data were collected from stools at the end of the study with 16S rRNA sequencing. We obtained sequence data from Arthur et al. \citep{Arthur2013} and generated open reference OTUs using the Quantitative Insights into Microbial Ecology (QIIME) \citep{Caporaso2010} version 1.9.1 at 97\% similarity level using the Greengenes 97\% reference dataset (release 13$\_$8). Chimeric sequences were detected and removed using QIIME. OTUs that had  0.005\% of the total number of sequences were excluded according to Bokulich and colleagues \citep{Bokulich2013}. Taxonomic assignment was done using the RDP (ribosomal database project) classifier \citep{Wang2007} through QIIME with confidence set to 50\%. There were 362 OTUs in total in the data sets after quality control and data cleaning. 40\% of the OTU RA data points were zero.

\begin{figure}[ht]
  \begin{center}
  \includegraphics[width =1\textwidth,angle=0]{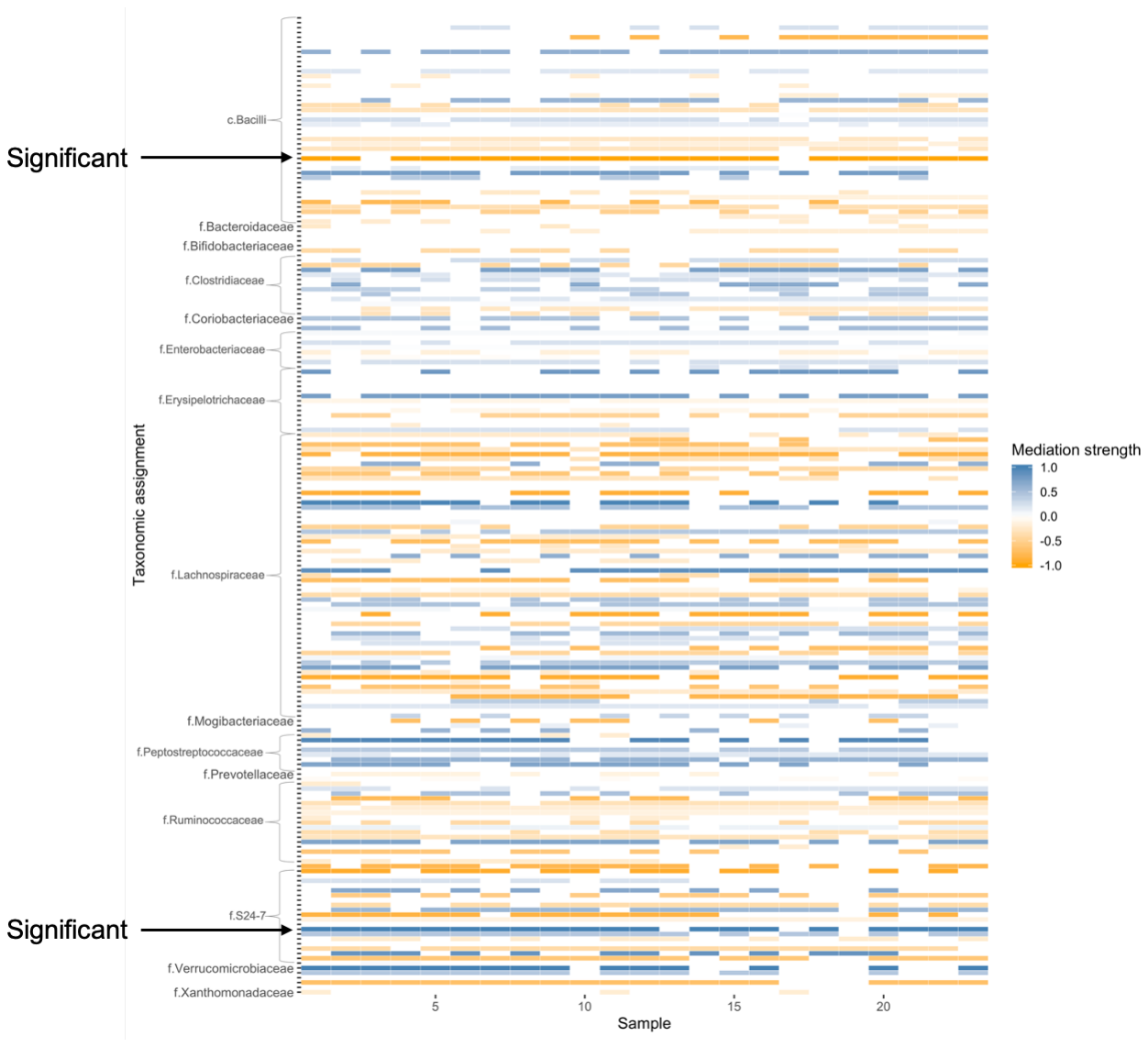}
  \end{center}
  \caption{Heatmap of mediation strength based on NIE$_1$ in VSL\#3 study. The mediation strength is measured by (1-p) where p is the unadjusted p-value. Negative sign indicates negative NIE$_1$. Taxonomic assignment is labeled on the vertical axis. Samples are labeled on the horizontal axis. Absence of an OTU in a sample is left blank in the heatmap.}
  \label{VSL_heatmap_NIE1}
\end{figure}

RA of each OTU was analyzed as a mediator variable using a ZIB distribution. The outcome variable in our analysis was dysplasia score (the higher the worse) which is a ordinal categorical variable measuring the abnormality of cell growth and it is treated as a continuous variable in the analysis because of its ordinal nature and its roughly bell-shaped density curve. The treatment variable is coded as 1/0 indicating VSL\#3/control. Again, the FDR approach was used for adjusting for multiple testing such that the targeted FDR is 20\% and the 95\% CI were calculated before adjustment.  NIE$_1$ of two OTUs were found to be statistically significant. One of the two OTUs was assigned to the family S24-7 under order Bacteroidales and the other one was assigned to class Bacilli. The estimates of NIE$_1$ were 0.27 (95\% CI: 0.1, 0.42) and -1.28 (95\% CI: -2.06, -0.49) respectively. The family S24-7 and class Bacilli found by our approach have also been reported to be related with colorectal cancer in the literature \citep{Peters2016,Braten2017}. To give a full picture of the mediation effects in this data set, a heatmap based on p-values was constructed (see Figure \ref{VSL_heatmap_NIE1}) to illustrate the NIE$_1$ of all OTUs. CCMM and IKT did not find any significant mediation effects of the OTUs. 

\section{Discussion}\label{sc:discu}
We developed an innovative marginal mediation modeling approach under the PO framework to analyze zero-inflated compositional mediators such as  microbiome. We showed that the mediation effect for zero-inflated mediators can be decomposed into two components of which the first is due to the change in the mediator over its positive domain and the second is due to the discrete binary change from zero to a non-zero status. These two components have different interpretations and are equally important for investigating causal mechanisms. The marginal model approach can also account for the compositional structure. When the point mass at zero (i.e., $\Delta$) is equal to zero for the mediator (i.e., the distribution is not zero-inflated), the model reduces to a marginal mediation model for data without zeros as described in Section \ref{sc:modelNo0}. Therefore, this approach can be also used for data sets after zero-valued data points are imputed with a positive number such as a Pseudocount (or after other normalization techniques are applied). R scripts for implementing the method are available upon request. 

This paper considered $X$ as a univariate variable and did not include covariates as potential confounders in the models. It is straightforward to adjust for a set of covariates using our approach. Let $C$ denote a vector of covariates or potential confounders. Then the NIE and NDE can be calculated at a specific value, $c$, of $C$ as $\text{NIE}=\E(Y_{x_2M_j(x_2)}-Y_{x_2M_j(x_1)}|C=c)$, $\text{NDE}=\E(Y_{x_2 M_j(x_1)}-Y_{x_1 M_j(x_1)}|C=c)$ and $\text{CDE}=\E(Y_{x_2 m}-Y_{x_1 m}|C=c)$. The value of $c$ can be taken as the mean value of the covariates similar to how least squares mean is calculated in regression models \citep{LSM1982}. CI can be obtained using the delta method or resampling methods. Decomposition of NIE follows the same procedure as shown in Section \ref{NIE_NDE}.

Misspecification of the mechanisms for observing zero-valued data points could have an impact on the model performance. This is similar to missing data issues where partial information is available on the missing data. It can be considered as missing not at random (MNAR) \citep{Little2014} because the probability of a data point being observed as zero depends on its true value. Besides the LOD mechanism in equation (\ref{eq:RAzero1}), another possible mechanism could be $\Pr(M_j^*=0|M_j,L)=\exp(-\eta M_jL)$ where $\eta>0$ and thus it is a decreasing function of $M_jL$, the SAA, such that smaller values of $M_jL$ are more likely to be observed as zero. Notice that the observed value $M_j^*$ is equal to zero with probability of one when $M_j=0$ which corresponds to the case that $M_j$ is truly zero. Model selection approaches such BIC or AIC can be used to choose different mechanisms. Although these mechanisms may not be perfect to account for MNAR, it can, to a large extent, alleviate the burden of not accounting for false zeros in the data at all. A future project has been planned to study the robustness of our model with respect to the mechanism for observing zeros using sensitivity analysis techniques.

\section{Appendix}

\subsection{Marginal association beween $Y$ and $M_j$ under equation (\ref{ge:1no0})}
Subject index $i$ is again suppressed in this section for simplicity. To obtain the marginal association beween $Y$ and $M_j$ under equation (\ref{ge:1no0}), we derive the expression for the conditional expectation $E_X(Y|M_j)$ which is the mean of $Y$ conditional on $M_j$ given $X$. By following basic principles of calculating conditional expectations, we have:
\begin{align}
&E_X(Y|M_j)=E_X\bigg(\sum_{k=1}^{K+1}\beta^kM_k+\beta^XX+\sum_{k=1}^{K+1}\beta^{kk}XM_k+\epsilon\bigg|M_j\bigg) \nonumber\\
&=\sum_{k=1}^{K+1}\beta^kE_X(M_k|M_j)+\beta^XX+\sum_{k=1}^{K+1}\beta^{kk}XE_X(M_k|M_j)+E_X\Big(\epsilon\Big|M_j\Big)\nonumber\\
&=\sum_{k=1}^{K+1}\beta^kE_X(M_k|M_j)+\beta^XX+\sum_{k=1}^{K+1}\beta^{kk}XE_X(M_k|M_j). \label{marginEqu}
\end{align}
Next we need to derive the expression for $E_X\Big(M_k\Big|M_j\Big)$ for all $k=1,\dots,K+1$ in the above equation. It is trivial to see that $E_X\Big(M_j\Big|M_j\Big)=M_j$. Let $M_{-j}$ denote the vector containing all but $M_j$ and thus $M_{-j}=(M_1,\dots,M_{j-1},M_{j+1},\dots,M_{K+1})^T$. Since $M$ has a Dirichlet distribution, the subcomposition $\frac{M_{-j}}{1-M_j}$ conditional on $M_j$ follows another Dirichlet distribution \citep{AitchJRSSB} with the mean parameters being $\bigg(\frac{\mu_1}{\sum_{k\ne j}\mu_k},\dots,\frac{\mu_{j-1}}{\sum_{k\ne j}\mu_k},\frac{\mu_{j+1}}{\sum_{k\ne j}\mu_k},\dots,\frac{\mu_{K+1}}{\sum_{k\ne j}\mu_k}\bigg)$ and the dispersion parameter being $\phi\sum_{k\ne j}\mu_k$. Thus, for any $M_k$ in the subvector $M_{-j}$, we have 
\begin{align*}
E_X\Big(M_k\Big|M_j\Big)&=E_X\bigg((1-M_j)\frac{M_k}{1-M_j}\bigg|M_j\bigg)\\
&=(1-M_j)E_X\bigg(\frac{M_k}{1-M_j}\bigg|M_j\bigg)\\
&=(1-M_j)\frac{\mu_k}{\sum_{l\ne j}\mu_l}.
\end{align*}
By plugging the above results into equation (\ref{marginEqu}), we have
\begin{align*}
&E_X(Y|M_j)=\sum_{k=1}^{K+1}\beta^kE_X(M_k|M_j)+\beta^XX+\sum_{k=1}^{K+1}\beta^{kk}XE_X(M_k|M_j) \\
&=\beta^jM_j+\sum_{k\ne j}\beta^k(1-M_j)\frac{\mu_k}{\sum_{l\ne j}\mu_l}+\beta^XX+\beta^{jj}XM_j+\sum_{k\ne j}\beta^{kk}X(1-M_j)\frac{\mu_k}{\sum_{l\ne j}\mu_l}\\
&=\beta_0^*+\beta_1^*M_j+\beta_2^*X+\beta_3^*XM_j,
\end{align*}
where 
\begin{align*}
\beta_0^*=\frac{\sum_{k\ne j}\beta^k\mu_k}{\sum_{l\ne j}\mu_l},\hspace{0.3cm} \beta_1^*=\beta^j-\beta_0^*,\hspace{0.3cm}
\beta_2^*=\beta^X+\frac{\sum_{k\ne j}\beta^{kk}\mu_k}{\sum_{l\ne j}\mu_l},\hspace{0.3cm}\text{and}\hspace{0.3cm}\beta_3^*=\beta^{jj}-\frac{\sum_{k\ne j}\beta^{kk}\mu_k}{\sum_{l\ne j}\mu_l}.
\end{align*}

\subsection{Multivariate delta method for obtaining 95\% CI of NIE$_1$, NIE$_2$, NDE and CDE}
Let $\zeta=(\beta_0,\beta_1,\beta_2,\beta_3,\beta_4,\beta_5,\delta,\alpha_0,\alpha_1,\gamma_0,\gamma_1)^\top$. The formulas for  NIE$_1$, NIE$_2$, NIE, NDE and CDE can be considered as functions of the full parameter vector $\zeta$. Let $f_1(\zeta)=\text{NIE}_1$ as derived in Section \ref{NIE_NDE} and thus $f_1(\hat\zeta)$ is the MLE of NIE$_1$ where $\hat\zeta$ is the MLE of $\zeta$. We first calculate the observed Fisher information matrix which can be calculated as $I_{obs}=-\frac{\partial^2 \ell}{\partial\zeta\partial\zeta^\top}|_{\zeta=\hat\zeta}$ where $\ell$ is the loglikelihood function in equation (\ref{logL}). By using the multivariate Delta method, we can calculate the variance of the estimator as follows:
\begin{align*}
\text{var}(\reallywidehat{\text{NIE}}_1)=\text{var}(f_1(\hat\zeta))&=\bigg(\frac{\partial f_1(\zeta)}{\partial\zeta}|_{\zeta=\hat\zeta}\bigg)^\top\text{var}(\hat\zeta)\bigg(\frac{\partial f_1(\zeta)}{\partial\zeta}|_{\zeta=\hat\zeta}\bigg)\\
&=\bigg(\frac{\partial f_1(\zeta)}{\partial\zeta}|_{\zeta=\hat\zeta}\bigg)^\top I_{obs}^{-1}\bigg(\frac{\partial f_1(\zeta)}{\partial\zeta}|_{\zeta=\hat\zeta}\bigg),
\end{align*}
where $\frac{\partial f_1(\zeta)}{\partial\zeta}=\bigg(\frac{\partial f_1(\zeta)}{\partial\beta_0},\frac{\partial f_1(\zeta)}{\partial\beta_1},\dots,\frac{\partial f_1(\zeta)}{\partial\gamma_1}\bigg)^\top$. Let $z_{0.025}$ denotes the 97.5th percentile of the standard normal distribution and the 95\% CI of NIE$_1$ can calculated as $\bigg(f_1(\hat\zeta)-z_{0.025}\sqrt{\text{var}(f_1(\hat\zeta))},f_1(\hat\zeta)+z_{0.025}\sqrt{\text{var}(f_1(\hat\zeta))}\bigg)$. The 95\% CI for NIE$_2$, NDE and CDE can be calculated similarly.

\section*{Acknowledgements}
Ths work was supported by U.S. NIH Grants R01GM123014, UH3OD023275, P01ES022832 and P20GM104416, and U.S. EPA grant RD 83544201.

\vspace{0.1cm}
\bibliography{../JabRef}

\begin{thebibliography}{}

\bibitem[\protect\astroncite{Aitchison}{1982}]{AitchJRSSB}
Aitchison, J. (1982).
\newblock The statistical analysis of compositional data.
\newblock {\em Journal of the Royal Statistical Society Series B-Statistical
  Methodology}.

\bibitem[\protect\astroncite{Arthur et~al.}{2013}]{Arthur2013}
Arthur, J.~C., Gharaibeh, R.~Z., Uronis, J.~M., Perez-Chanona, E., Sha, W.,
  Tomkovich, S., M{\"u}hlbauer, M., Fodor, A.~A., and Jobin, C. (2013).
\newblock Vsl\# 3 probiotic modifies mucosal microbial composition but does not
  reduce colitis-associated colorectal cancer.
\newblock {\em Scientific reports}, 3:2868.

\bibitem[\protect\astroncite{Baron and Kenny}{1986}]{BaronKenny}
Baron, R.~M. and Kenny, D.~A. (1986).
\newblock The moderator-mediator variable distinction in social psychological
  research: Conceptual, strategic and statistical considerations.
\newblock {\em Journal of Personality and Social Psychology}, 51:1173--1182.

\bibitem[\protect\astroncite{Belkaid and Hand}{2014}]{Belkaid2014}
Belkaid, Y. and Hand, T.~W. (2014).
\newblock Role of the microbiota in immunity and inflammation.
\newblock {\em Cell}, 157:121--141.

\bibitem[\protect\astroncite{Benjamini and Hochberg}{1995}]{Benjamini1995}
Benjamini, Y. and Hochberg, Y. (1995).
\newblock {Controlling the false discovery rate: A Practical and powerful
  approach to multiple testing}.
\newblock {\em J. Roy. Statist. Soc.}, 57:289--300.

\bibitem[\protect\astroncite{Bokulich et~al.}{2013}]{Bokulich2013}
Bokulich, N.~A., Subramanian, S., Faith, J.~J., Gevers, D., Gordon, J.~I.,
  Knight, R., Mills, D.~A., and Caporaso, J.~G. (2013).
\newblock Quality-filtering vastly improves diversity estimates from illumina
  amplicon sequencing.
\newblock {\em Nature methods}, 10(1):57.

\bibitem[\protect\astroncite{Br{\aa}ten et~al.}{2017}]{Braten2017}
Br{\aa}ten, L.~S., S{\o}dring, M., Paulsen, J.~E., Snipen, L.~G., and Rudi, K.
  (2017).
\newblock Cecal microbiota association with tumor load in a colorectal cancer
  mouse model.
\newblock {\em Microbial ecology in health and disease}, 28(1):1352433.

\bibitem[\protect\astroncite{Caporaso et~al.}{2010}]{Caporaso2010}
Caporaso, J.~G., Kuczynski, J., Stombaugh, J., Bittinger, K., Bushman, F.~D.,
  Costello, E.~K., Fierer, N., Pena, A.~G., Goodrich, J.~K., Gordon, J.~I.,
  et~al. (2010).
\newblock Qiime allows analysis of high-throughput community sequencing data.
\newblock {\em Nature methods}, 7(5):335.

\bibitem[\protect\astroncite{Chai et~al.}{2018}]{Chai2018}
Chai, H., Jiang, H., Lin, L., and Liu, L. (2018).
\newblock A marginalized two-part beta regression model for microbiome
  compositional data.
\newblock {\em PLoS computational biology}, 14(7):e1006329.

\bibitem[\protect\astroncite{Chen and Li}{2016}]{Chen2016}
Chen, E.~Z. and Li, H. (2016).
\newblock A two-part mixed-effects model for analyzing longitudinal microbiome
  compositional data.
\newblock {\em Bioinformatics (Oxford, England)}, 32:2611--2617.

\bibitem[\protect\astroncite{Cribari-Neto and Zeileis}{2010}]{betaReg}
Cribari-Neto, F. and Zeileis, A. (2010).
\newblock Beta regression in {R}.
\newblock {\em Journal of Statistical Software}, 34:24848.

\bibitem[\protect\astroncite{Dalrymple et~al.}{2003}]{Dalrymple2003}
Dalrymple, M.~L., Hudson, I.~L., and Ford, R. P.~K. (2003).
\newblock Finite mixture, zero-inflated {P}oisson and hurdle models with
  application to {SIDS}.
\newblock {\em Computational Statistics \& Data Analysis}, 41(3-4):491--504.

\bibitem[\protect\astroncite{Efron and Tibshirani}{1986}]{efron1986}
Efron, B. and Tibshirani, R. (1986).
\newblock Bootstrap methods for standard errors, confidence intervals, and
  other measures of statistical accuracy.
\newblock {\em Statistical Science}, 1(1):54--75.

\bibitem[\protect\astroncite{Ferrari and Cribari-Neto}{2004}]{betaDist}
Ferrari, S. and Cribari-Neto, F. (2004).
\newblock Beta regression for modelling rates and proportions.
\newblock {\em Journal of Applied Statistics}, 31:799--815.

\bibitem[\protect\astroncite{Gianola}{1982}]{LSM1982}
Gianola, D. (1982).
\newblock Least-squares means vs population marginal means.
\newblock {\em American Statistician}, 36(1):65--66.

\bibitem[\protect\astroncite{Gionchetti et~al.}{2000}]{Gionchetti2000}
Gionchetti, P., Rizzello, F., Venturi, A., Brigidi, P., Matteuzzi, D.,
  Bazzocchi, G., Poggioli, G., Miglioli, M., and Campieri, M. (2000).
\newblock Oral bacteriotherapy as maintenance treatment in patients with
  chronic pouchitis: a double-blind, placebo-controlled trial.
\newblock {\em Gastroenterology}, 119(2):305--309.

\bibitem[\protect\astroncite{Imai et~al.}{2010}]{Imai}
Imai, K., Keele, L., and Tingley, D. (2010).
\newblock A general approach to causal mediation analysis.
\newblock {\em Psychological Methods}, 15:309--334.

\bibitem[\protect\astroncite{Jin et~al.}{2019}]{Jin2019}
Jin, C., Lagoudas, G.~K., Zhao, C., Bullman, S., Bhutkar, A., Hu, B., Ameh, S.,
  Sandel, D., Liang, X.~S., Mazzilli, S., Whary, M.~T., Meyerson, M., Germain,
  R., Blainey, P.~C., Fox, J.~G., and Jacks, T. (2019).
\newblock Commensal microbiota promote lung cancer development via gammadelta t
  cells.
\newblock {\em Cell}, 176:998--1013.e16.

\bibitem[\protect\astroncite{Lange et~al.}{2017}]{Lange2017}
Lange, T., Hansen, K.~W., Sørensen, R., and Galatius, S. (2017).
\newblock Applied mediation analyses: a review and tutorial.
\newblock {\em Epidemiology and health}, 39:e2017035.

\bibitem[\protect\astroncite{Li}{2018}]{hongzhe18review}
Li, H. (2018).
\newblock Statistical and computational methods in microbiome and metagenomics.
\newblock {\em Handbook in Statistical Genomics}.

\bibitem[\protect\astroncite{Little and Rubin}{2014}]{Little2014}
Little, R.~J. and Rubin, D.~B. (2014).
\newblock {\em Statistical analysis with missing data}, volume 333.
\newblock John Wiley \& Sons.

\bibitem[\protect\astroncite{MacKinnon}{2008}]{MacKin08}
MacKinnon, D.~P. (2008).
\newblock {\em Introduction to statistical mediation analysis}.
\newblock New York: Erlbaum.

\bibitem[\protect\astroncite{MacKinnon et~al.}{2007}]{MacKinnon2007}
MacKinnon, D.~P., Fairchild, A.~J., and Fritz, M.~S. (2007).
\newblock Mediation analysis.
\newblock {\em Annual review of psychology}, 58:593--614.

\bibitem[\protect\astroncite{Madsen et~al.}{2001}]{Madsen2001}
Madsen, K., Cornish, A., Soper, P., McKaigney, C., Jijon, H., Yachimec, C.,
  Doyle, J., Jewell, L., and De~Simone, C. (2001).
\newblock Probiotic bacteria enhance murine and human intestinal epithelial
  barrier function.
\newblock {\em Gastroenterology}, 121(3):580--591.

\bibitem[\protect\astroncite{Martinez and Bartholomew}{2017}]{Hmean}
Martinez, M.~N. and Bartholomew, M.~J. (2017).
\newblock What does it "mean"? a review of interpreting and calculating
  different types of means and standard deviations.
\newblock {\em Pharmaceutics}, 9(2).

\bibitem[\protect\astroncite{Pagnini et~al.}{2010}]{Pagnini2010}
Pagnini, C., Saeed, R., Bamias, G., Arseneau, K.~O., Pizarro, T.~T., and
  Cominelli, F. (2010).
\newblock Probiotics promote gut health through stimulation of epithelial
  innate immunity.
\newblock {\em Proceedings of the national academy of sciences},
  107(1):454--459.

\bibitem[\protect\astroncite{Peters et~al.}{2016}]{Peters2016}
Peters, B.~A., Dominianni, C., Shapiro, J.~A., Church, T.~R., Wu, J., Miller,
  G., Yuen, E., Freiman, H., Lustbader, I., Salik, J., et~al. (2016).
\newblock The gut microbiota in conventional and serrated precursors of
  colorectal cancer.
\newblock {\em Microbiome}, 4(1):69.

\bibitem[\protect\astroncite{Sohn and Li}{2019}]{Sohn2019}
Sohn, M.~B. and Li, H. (2019).
\newblock Compositional mediation analysis for microbiome studies.
\newblock {\em The Annals of Applied Statistics}, 13(1):661--681.

\bibitem[\protect\astroncite{Sood et~al.}{2009}]{Sood2009}
Sood, A., Midha, V., Makharia, G.~K., Ahuja, V., Singal, D., Goswami, P., and
  Tandon, R.~K. (2009).
\newblock The probiotic preparation, vsl\# 3 induces remission in patients with
  mild-to-moderately active ulcerative colitis.
\newblock {\em Clinical Gastroenterology and Hepatology}, 7(11):1202--1209.

\bibitem[\protect\astroncite{Tanoue et~al.}{2019}]{Tanoue2019}
Tanoue, T., Morita, S., Plichta, D.~R., Skelly, A.~N., Suda, W., Sugiura, Y.,
  Narushima, S., Vlamakis, H., Motoo, I., Sugita, K., Shiota, A., Takeshita,
  K., Yasuma-Mitobe, K., Riethmacher, D., Kaisho, T., Norman, J.~M., Mucida,
  D., Suematsu, M., Yaguchi, T., Bucci, V., Inoue, T., Kawakami, Y., Olle, B.,
  Roberts, B., Hattori, M., Xavier, R.~J., Atarashi, K., and Honda, K. (2019).
\newblock A defined commensal consortium elicits cd8 t cells and anti-cancer
  immunity.
\newblock {\em Nature}, 565:600--605.

\bibitem[\protect\astroncite{Terhorst}{1986}]{stiel}
Terhorst, H.~J. (1986).
\newblock On stieltjes integration in euclidean-space.
\newblock {\em Journal of Mathematical Analysis and Applications},
  114(1):57--74.

\bibitem[\protect\astroncite{Tingley et~al.}{2017}]{mediationRpackage}
Tingley, D., Yamamoto, T., Hirose, K., Keele, L., and Imai, K. (2017).
\newblock mediation: R package for causal mediation analysis.
\newblock {\em
  https://cran.r-project.org/web/packages/mediation/vignettes/mediation.pdf}.

\bibitem[\protect\astroncite{VanderWeele}{2009}]{Vander09}
VanderWeele, T.~J. (2009).
\newblock Marginal structural models for the estimation of direct and indirect
  effects.
\newblock {\em Epidemiology}, 20:18--26.

\bibitem[\protect\astroncite{VanderWeele}{2015}]{Vander15}
VanderWeele, T.~J. (2015).
\newblock {\em Explanation in Causal Inference: Methods for Mediation and
  Interaction}.
\newblock New York: Oxford Univ. Press.

\bibitem[\protect\astroncite{VanderWeele}{2016}]{Vander16}
VanderWeele, T.~J. (2016).
\newblock Mediation analysis: A practitioner's guide.
\newblock {\em Annu Rev Public Health}, 37:17--32.

\bibitem[\protect\astroncite{Wang et~al.}{2019a}]{chanWang2019}
Wang, C., Hu, J., Blaser, M.~J., and Li, H. (2019a).
\newblock Estimating and testing the microbial causal mediation effect with
  high-dimensional and compositional microbiome data.
\newblock {\em Bioinformatics}.

\bibitem[\protect\astroncite{Wang et~al.}{2007}]{Wang2007}
Wang, Q., Garrity, G.~M., Tiedje, J.~M., and Cole, J.~R. (2007).
\newblock Naive bayesian classifier for rapid assignment of rrna sequences into
  the new bacterial taxonomy.
\newblock {\em Appl. Environ. Microbiol.}, 73(16):5261--5267.

\bibitem[\protect\astroncite{Wang et~al.}{2019b}]{Wang2019a}
Wang, X., Sun, G., Feng, T., Zhang, J., Huang, X., Wang, T., Xie, Z., Chu, X.,
  Yang, J., Wang, H., Chang, S., Gong, Y., Ruan, L., Zhang, G., Yan, S., Lian,
  W., Du, C., Yang, D., Zhang, Q., Lin, F., Liu, J., Zhang, H., Ge, C., Xiao,
  S., Ding, J., and Geng, M. (2019b).
\newblock Sodium oligomannate therapeutically remodels gut microbiota and
  suppresses gut bacterial amino acids-shaped neuroinflammation to inhibit
  alzheimer's disease progression.
\newblock {\em Cell research}, 29:787--803.

\bibitem[\protect\astroncite{Zhang et~al.}{2019}]{LeiLiu2019}
Zhang, H., Chen, J., Li, Z., and Liu, L. (2019).
\newblock Testing for mediation effect with application to human microbiome
  data.
\newblock {\em Statistics in Biosciences}, In press.

\end{thebibliography}
\bibliographystyle{apa}

\end{document}